# Hydration of a side-chain-free n-type semiconducting ladder polymer driven by electrochemical doping

Jiajie Guo,[1,2] Lucas Q. Flagg,[3] Duyen K. Tran,[4] Shinya E. Chen,[1,2] Ruipeng Li,[5] Nagesh B. Kolhe,[4] Rajiv Giridharagopal,[2] Samson A. Jenekhe,[4] Lee J. Richter,[3] and David S. Ginger[2,6]*

[1] Molecular Engineering and Sciences Institute, University of Washington, Seattle, WA, US 98195

[2] Department of Chemistry, University of Washington, Seattle, WA, USA 98195

[3] Materials Science and Engineering Division, National Institute of Standards and Technology, Gaithersburg, Maryland, USA 20899

[4] Department of Chemical Engineering, University of Washington, Seattle, WA, USA 98195

[5] National Synchrotron Light Source II, Brookhaven National Laboratory, Upton, NY, USA 11973

[6] Physical Sciences Division, Physical and Computational Sciences Directorate, Pacific Northwest National Laboratory, Richland, WA, USA 99352

# ABSTRACT

We study the organic electrochemical transistor**s** (OECTs) performance of the ladder polymer, poly(benzimidazobenzophenanthroline) (BBL) in an attempt to better understand how an apparently hydrophobic side-chain-free polymer is able to operate as an OECT with favorable redox kinetics in an aqueous environment. We examine two BBLs of different molecular masses from different sources. Both BBLs show significant film swelling during the initial reduction step. By combining electrochemical quartz crystal microbalance (eQCM) gravimetry, *in-operando* atomic force microscopy (AFM), and both *ex-situ* and *in-operando* grazing incidence wide-angle x-ray scattering (GIWAXS), we provide a detailed structural picture of the electrochemical charge injection process in BBL in the absence of any hydrophilic side-chains. Compared with *ex-situ* measurements, *in-operando* GIWAXS shows both more swelling upon electrochemical doping than has previously been recognized, and less contraction upon dedoping. The data show that BBL films undergo an irreversible hydration driven by the initial electrochemical doping cycle with significant water retention and lamellar expansion that persists across subsequent oxidation/reduction cycles. This swelling creates a hydrophilic environment that facilitates the subsequent fast hydrated ion transport in the absence of the hydrophilic side-chains used in many other polymer systems. Due to its rigid ladder backbone and absence of hydrophilic side-chains, the primary BBL water uptake does not significantly degrade the crystalline order, and the original dehydrated, unswelled state can be recovered after drying. The combination of doping induced hydrophilicity and robust crystalline order leads to efficient ionic transport and good stability.

**TOC graphic**

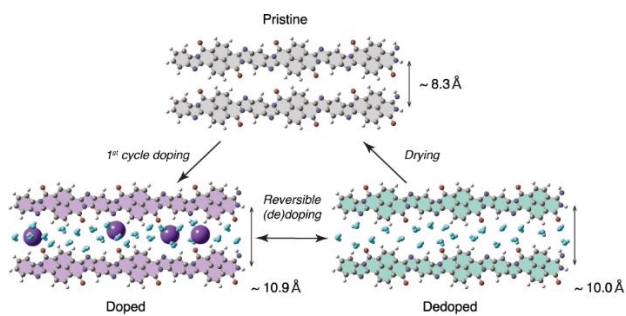

# INTRODUCTION

Conjugated polymers showing mixed ionic-electronic transport,[1-4] more generally organic mixed-ionic electronic conductors (OMIECs), have emerged in the past few decades as promising materials for applications in bioelectronics,[5-8] energy storage,[9-10] logic circuit elements,[11-12] and neuromorphic computing.[13-15] Organic electrochemical transistors (OECTs) are widely used as a model testbed to study mixed ionic-electronic transport in OMIECs.[16-18] Unlike conventional organic field-effect transistors (OFETs), an OECT's source-drain current ($I_D$) is modulated by the gate ($V_G$) through voltage-driven redox processes and associated ion uptake from the electrolyte into the entire volume of the semiconductor channel, resulting in bulk doping/dedoping. Hence, the efficiency of this modulation, described by the transconductance ($g_m \equiv \partial I_D/\partial V_G$), is governed by the carrier transport mobility ($\mu$) and the volumetric capacitance ($C^*$) in the channel.[19] Tremendous progress has been achieved in developing high-performance p-type (hole transport) materials, with $\mu C^*$ value on the order of hundreds of F cm$^{-1}$ V$^{-1}$ s$^{-1}$.[20-21] However, n-type organic materials have generally lagged behind.[22-26] The lack of n-type materials limits applications such as biosensors involving cations, complementary circuit logic, as well as electrochemical energy storage.

In recent years, considerable effort has been devoted to the design and synthesis of novel n-type OMIECs, mainly based on donor-acceptor (D–A) copolymers.[23] The most popular molecular design strategy is to use hydrophilic side-chains to help ionic transport in aqueous electrolyte, for both p-type[27-30] and n-type[4, 22, 30-31] materials. However, studies suggest that excessive water uptake harms the electronic mobility and stability when adding hydrophilic glycol side-chains.[4, 32-33] In this context, the polymer poly(benzimidazobenzophenanthroline) (BBL, structure shown in **Figure 1A**), stands out as a notable exception that functions without glycol side-chains.[11, 34-35] Indeed, its electron mobility in OECTs has been reported as an order of magnitude higher than D–A based n-type polymers, presumably due to the planar, rigid BBL backbone.[6] Recently, Fabiano *et al.* demonstrated that increasing the molecular mass yields an improvement of electronic mobility.[34] Furthermore, Inal *et al.* reported that the side-chain-free BBL accommodates a higher density of cations and water, compared to the more conventional naphthalenediimide(NDI)- based glycolated polymer, P-90.[35] As a result, BBL shows a record high volumetric capacitance in n-type OMIECs and somehow overcomes the ubiquitous $\mu\text{-}C^*$ trade-off, that commonly plagues p-type materials. Nevertheless, a detailed understanding of how water and cations are able to enter such a hydrophobic film is still lacking.

Here, we explore the electrochemical doping process using a wide range of complementary analytical methods on both commercial low molecular mass BBL (**BBL_L**) and in-house synthesized high molecular mass BBL (**BBL_H**). In both versions of BBLs, we observe an irreversible water mass uptake during the first electrochemical doping step via electrochemical quartz crystal microbalance (eQCM), which agrees with the *in-operando* AFM film thickness change. Furthermore, we probe the nanoscale structure change by electrochemical strain microscopy (ESM) *in-operando*. Finally, we characterize the BBL crystalline structure changes using both *ex-situ* and *in-operando* grazing incidence wide-angle scattering (GIWAXS). By combining these complementary results, we propose a unique film hydration behavior in the side-chain-free ladder polymer BBL during the electrochemical reduction. In absence of the electrochemical bias, the hydrophobic BBL film shows resistance to water uptake in both crystalline and amorphous regions. However, during the initial electrochemical doping, BBL undergoes an ≈30 % lamellar expansion upon water injection, which remains persistent across the subsequent doping/dedoping cycles. We propose this retention of water between the lamellar structure creates a hydrophilic microenvironment for fast ion movement. Importantly, this expansion is not observed in *ex-situ* samples, which suggests the water uptake is reversible upon film drying. These results provide guidance for designing new OMIEC polymers that can overcome the tradeoff between ionic and electronic transport and achieve a high $\mu C^*$.

# RESULTS AND DISCUSSION

**Figure 1A** shows the chemical structure of the BBL polymer we studied. In this report, we studied two different BBLs, one synthesized in house by the Jenekhe group (**BBL$_H$**, $M_v$ = 60.5 kDa),[2] and one purchased commercially from Sigma-Aldrich (**BBL$_L$**, $M_v$ = 6.15 kDa). The molecular mass difference was also observed by dynamic light scattering (DLS) as shown in **Figure S2**.

First, we evaluated the OECTs performance of both versions of BBLs. In these OECT measurements, we used a Ag/AgCl pellet as the gate electrode, and 100 mmol/L degassed aqueous potassium chloride (KCl) as the electrolyte. **Figure 1B–C** show a typical transfer curve along with transconductance, $g_m$, and output curves for both **BBL$_L$** and **BBL$_H$**, respectively. The drain current and transconductance of the **BBL$_H$** OECTs are consistently two orders of magnitude higher than the **BBL$_L$** OECTs, and the peak transconductance, $g_{m,\,peak}$, appears at a lower gate voltage for the **BBL$_H$**. Also, the threshold voltage is slightly more negative (≈100 mV) for the **BBL$_H$**. (**Figure S3**) A similar threshold voltage shift was observed in other BBL OECT studies.[34] The shift in both peak transconductance and threshold voltage suggests that the **BBL$_H$** is energetically easier to be electrochemically doped in KCl. This difference also agrees with the observed onset voltage difference in cyclic voltammetry (CV) measurements of two BBLs. (**Figure S4**)

Compared to OFETs, OECTs have the advantage of volumetric capacitance, wherein the entire bulk of the active OMIEC film, instead of the surface, can accommodate ions and electronic carriers.[36] After rewriting the conventional field-effect transistor equation to replace the gate capacitance, $C$, with the effective volumetric capacitance, $C^*$, the transconductance of OECTs in the saturation regime, is given by[19]:

$$g_m = \frac{\partial I_D}{\partial V_G} = \frac{Wd}{L}\mu C^*|V_G - V_T| \qquad \textbf{(Equation 1)}$$

where $W$, $d$, and $L$ are the channel width, thickness, and length, respectively, $\mu$ is the OECT channel carrier mobility, $C^*$ is the OECT channel volumetric capacitance, $V_G$ is the operating gate voltage, and $V_T$ is the threshold voltage. Following **Eq. 1**, we extracted the $\mu C^*$ products of the two different BBLs from a linear regression between peak transconductance $g_{m,\text{peak}}$ and $\left(\frac{Wd}{L}\right)|V_G - V_T|$ obtained from a series of devices with different $Wd/L$ ratios. **Figure 1D** shows that our BBL OECT data is well fit to **Eq. 1**, with $\mu C^*$ increasing more than 30 times going from **BBL$_L$** ((0.28 ± 0.01) F cm$^{-1}$ V$^{-1}$ s$^{-1}$) to **BBL$_H$** ((9.27 ± 0.03) F cm$^{-1}$ V$^{-1}$ s$^{-1}$).

For an OECT, especially in the applications of sensors and complementary circuits, response time is also a crucial property, which relates to the ion migration from the electrolyte to the active polymer film. With such high volumetric capacitance, the BBL transistors can still be switched ON/OFF in tens of milliseconds, which suggests an unexpectedly fast ion transport in such hydrophobic films. (**Figure S5**) In addition, we also examine the stability of the superior **BBL$_H$** devices with continuous pulse gate potential over 100 cycles, and find they retain ≈97 % of the original performance. (**Figure S6**)

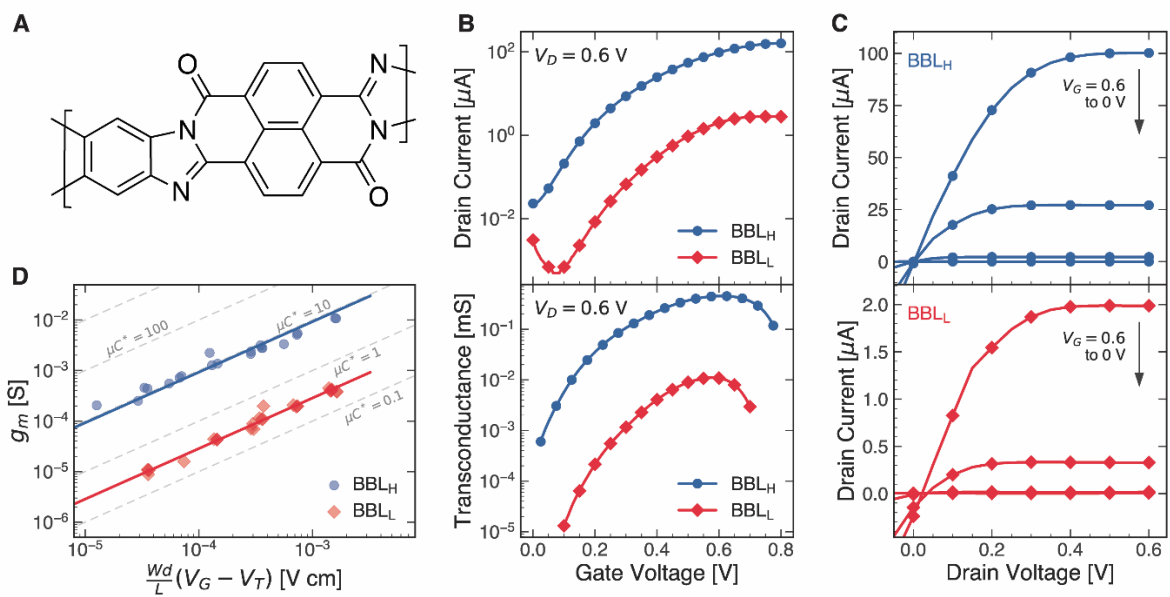

**Figure 1.** A) The chemical structure of BBL. B) Typical OECT transfer curves with calculated transconductances and C) output curves for both **BBL$_L$** and **BBL$_H$** OECTs in 100 mmol/L KCl. (channel width = 100 µm, channel length = 10 µm, channel thickness ≈120 nm) D) The peak transconductance $g_{m,\text{peak}}$ of BBL OECTs as a function of operating conditions and channel geometry. Each data point represents one OECT measurement. The linear slope of the data determines the OECT figure of merit $\mu C^*$. Dashed lines denote constant µC$^*$ product.

To further study the contribution of the carrier mobility ($\mu$) and volumetric capacitance ($C^*$) separately, we next measured the volumetric capacitance using electrochemical impedance spectroscopy (EIS). We fit the EIS spectra using a modified Randles circuit: $R_s$(CPE//$R_{ct}$), where $R_s$ is the electrolyte resistance, CPE is a constant phase element, $R_{ct}$ is the charge transfer resistance. (**Figure S7** and **Figure S8**) The constant phase element (CPE) can be interpreted as arising from the distribution of counterion-polaron pairs in the bulky polymer films.[37] We observed an improvement in the volumetric capacitance $C^*$ with the **BBL$_H$**, which suggests that the **BBL$_H$** film has a higher density of ion-accessible sites. (**Figure S8**) Taking the threshold voltage shift into account to get a fair comparison, the **BBL$_H$** and **BBL$_L$** films show average $C^*$ values of (1007.1 ± 172.8) F/cm$^3$ and (539.8 ± 85.8) F/cm$^3$ at 0.7 V and 0.8 V, respectively. The higher $C^*$ of **BBL$_H$** is also consistent with its higher CV current level, indicating more electronic charge injection. (**Figure S4**) Based on the $\mu C^*$ from the transistor measurements and the $C^*$ from EIS, we determined the electron mobilities of both BBLs. We calculate that **BBL$_H$** (9.2 ± 1.6 × 10$^{-3}$ cm$^2$ V$^{-1}$ s$^{-1}$) has a mobility that is over an order of magnitude higher than that for **BBL$_L$** (0.52 ± 0.08 × 10$^{-3}$ cm$^2$ V$^{-1}$ s$^{-1}$). Notably, the mobility we measure for **BBL$_L$** is in good agreement with the mobility of 10$^{-3}$ cm$^2$ V$^{-1}$ s$^{-1}$ previously reported for both solid-state OFETs[38] and wet OECTs[11] with the same commercial **BBL$_L$**. The difference in mobility implies that the **BBL$_H$** with a higher molecular mass, results in an enhanced electron transport process.[34] In conjugated polymers, the relationship between mobility, molecular mass, and morphology is complicated,[39-41] but the enhanced mobility of BBL follows the trend seen for many other materials.[34, 42-43] A comparison of the OECT performance of two BBLs are summarized in **Table 1**. We note that, in addition to higher mobility and volumetric capacitance, **BBL$_H$** shows faster switching times both turning ON and turning OFF processes.

**Table 1.** Summary of BBL OECTs performance.

| | ON/OFF ratio [†] | $V_T$ [†] [V] | $g_m$ [†] [mS] | $\tau_{on}$ [†] [ms] | $\tau_{off}$ [†] [ms] | $\mu C^{*a}$ [F cm$^{-1}$ V$^{-1}$ s$^{-1}$] | $C^{*b}$ [F cm$^{-3}$] | $\mu^c$ [cm$^2$ V$^{-1}$ s$^{-1}$] |
|---|---|---|---|---|---|---|---|---|
| **BBL$_H$** | $1.59 \times 10^4$ | 0.17 | 10.8 | 80.3 | 6.5 | $9.27 \pm 0.03$ | $1007.1 \pm 172.8$ | $9.2 \pm 1.6 \times 10^{-3}$ |
| **BBL$_L$** | $1.23 \times 10^4$ | 0.28 | 0.41 | 142 | 181 | $0.28 \pm 0.01$ | $539.8 \pm 85.8$ | $0.52 \pm 0.08 \times 10^{-3}$ |

All error bars represent standard error of the mean.

[†] Average data extracted from a channel with a width of 4000 μm and a length of 100 μm. [a] Extracted from Eq.1 with peak $g_{m,\,peak}$ (shown in **Figure 1D**); [b] Obtained from EIS measurements for various electrode areas; [c] Obtained via dividing μC* by C*.

To further understand the difference in volumetric capacitance values, we used spectroelectrochemistry and coulometry to characterize the polaron/bipolaron formation. The pristine films show similar UV−vis absorption spectra, while the π–π$^*$ transition peak (≈560 nm) redshifts as the molecular mass increases. (**Figure S9**) This peak shift suggests that the **BBL$_H$** has a longer effective conjugation length, which agrees with the electronic mobility enhancement in **BBL$_H$**.[44-46] **Figure S10–Figure S12** show the spectra change for the two polymers upon the electrochemical doping process under different bias. When a positive bias is applied to the gate, the BBL films are electrochemically reduced. The π–π* transition of the ground state intensity first quenched at a low voltage (below 0.5 V) in the range of 450 nm to 650 nm, accompanied by two isosbestic points at 435 nm and 670 nm. Simultaneously, a broad absorption peak grows in the range of 700 nm to 900 nm, indicating a transition from the neutral BBL into a charged BBL.[35, 38] Following previous studies, we assign the peak around 880 nm (labeled as B) as a polaron, and the peak around 460 nm (labeled as A) as a bipolaron.[2] At the same potential, the polaron and bipolaron densities formed in the **BBL$_L$** are always lower than the **BBL$_H$**, which is consistent with the smaller volumetric capacitance value of **BBL$_L$**. (**Figure S11**) We also cross-checked these doping densities determined via spectroelectrochemistry by using coulometry. We calculated the number of electrons injected during the reduction by integrating the pre-doping current prior to the UV–Vis spectra measurements and normalized it to the film volume. With this coulometry approach **BBL$_H$** shows an ≈2-fold higher doping level than the **BBL$_L$**, reaching 2 electrons per repeat unit at 0.9 V shown in **Figure S11**. This result agrees with our volumetric capacitance data from the EIS fitting. Although the spectroelectrochemistry measurement was carried out in a degassed electrolyte, the whole setup is exposed ambient air during measurements. Since the oxygen reduction reaction (ORR) may then contribute to the current,[35, 47] we note that the electron densities reported here via coulometry might be slightly overestimated due to ambient conditions.

When interacting with aqueous solutions, the swelling of conjugated polymer films is commonly divided into two classifications: 1) *passive swelling*: the hydration of a dry film by the electrolyte without applied potential; 2) *active swelling*: water injection that occurs during active redox processes. Previous work has shown that polymers with a large fraction of glycol side-chains can undergo significant water uptake and retention, even in the absence of applied potential.[4, 17, 20, 32] On the other hand, more hydrophobic polymers with few hydrophilic side-chains, which can behave like OECTs, often only exhibit active swelling.[4, 32, 48] As mentioned, BBL lacks such side-chains, therefore the conventional description of aqueous operation does not apply. We therefore need to understand what physical processes allow BBL to operate as an OECT as well as what doping mechanisms are affected by molecular mass.

In this context, we further investigate the film swelling of two BBLs during the electrochemical doping process using gravimetric measurements with an electrochemical quartz crystal microbalance (eQCM). We first probe the *passive* swelling, which relates to the film hydration after contact with the aqueous solution without bias. The mass of BBL film is determined by the Sauerbrey equation both in air and aqueous solution since BBL films are sufficiently thin and exhibit a low degree of swelling (less than 0.1% change of the resonance frequency) such that viscoelastic corrections are not needed.[35, 49-51] It is not surprising that without any hydrophilic side-chains, both BBL films show minimal swelling, whereas other n-type polymers with alkyl-glycol side-chains show massive swelling, up to ≈100 %.[4]

We then characterized *active* swelling. In previous p-type polymers studies, it is known that the ions are injected into (or removed from) the film surrounded by a hydration shell with some number of water molecules, and perhaps additional bulk water.[4] With the simultaneous recording of current, we calculate the mass of the injected cations by integrating the charges, assuming these are the only charged species getting into the film. We then calculate the water mass by subtracting the cation mass from the total mass, thus providing an estimate of the number of waters dragged into the film during cation injection. Interestingly, in contrast to other OMIEC polymers with alkyl-glycol side-chains showing continuous retention of mass accumulation during active swelling,[28, 52] the mass change in BBL shows two distinct phases. During the very first cycle from neat wet films, both molecular mass versions BBL films show a dramatic mass retention (82 % for **BBL$_H$** and 74 % for **BBL$_L$**) that is not reversible upon subsequent cycles (**Figure 2A**). However, the mass of ions remaining in the dedoped films after the first cycle is only about 3.4 % for **BBL$_H$** and 2.8 % for **BBL$_L$**, respectively. Thus, we propose that the electrochemically irreversible mass gain is mainly caused by water molecules, not the cations. Notably however, this electrochemically-irreversible first step water uptake could be regenerated on the same crystal after drying under a vacuum, suggesting that it does not permanently change the film morphology.

During subsequent electrochemical cycles, the film shows a reversible mass change with a stable baseline (**Figure 2B**). As expected, **Figure S14** shows that upon electrochemical doping the **BBL$_H$** incorporates more charges, along with more mass uptake, which agrees with the volumetric capacitance difference. In addition, **BBL$_H$** not only takes up more cations (16 % for **BBL$_H$** and 7.7 % for **BBL$_L$**), but also more water molecules compared to the **BBL$_L$**. We convert this mass into the number of water molecules per cation. Considering the onset shift between two BBLs, we still observe that more water molecules are dragged into the film per cation in the **BBL$_H$** (20 to 30 waters) than in the **BBL$_L$** (5 to 7 waters) at a doped state. According to previous studies of the hydration shell in aqueous solutions,[53-54] a potassium ion normally has a shell of ≈6 waters. From this we conclude that the mass uptake in **BBL$_H$** must involve free/bulk water uptake upon doping, apart from the waters in the cation hydration shells.

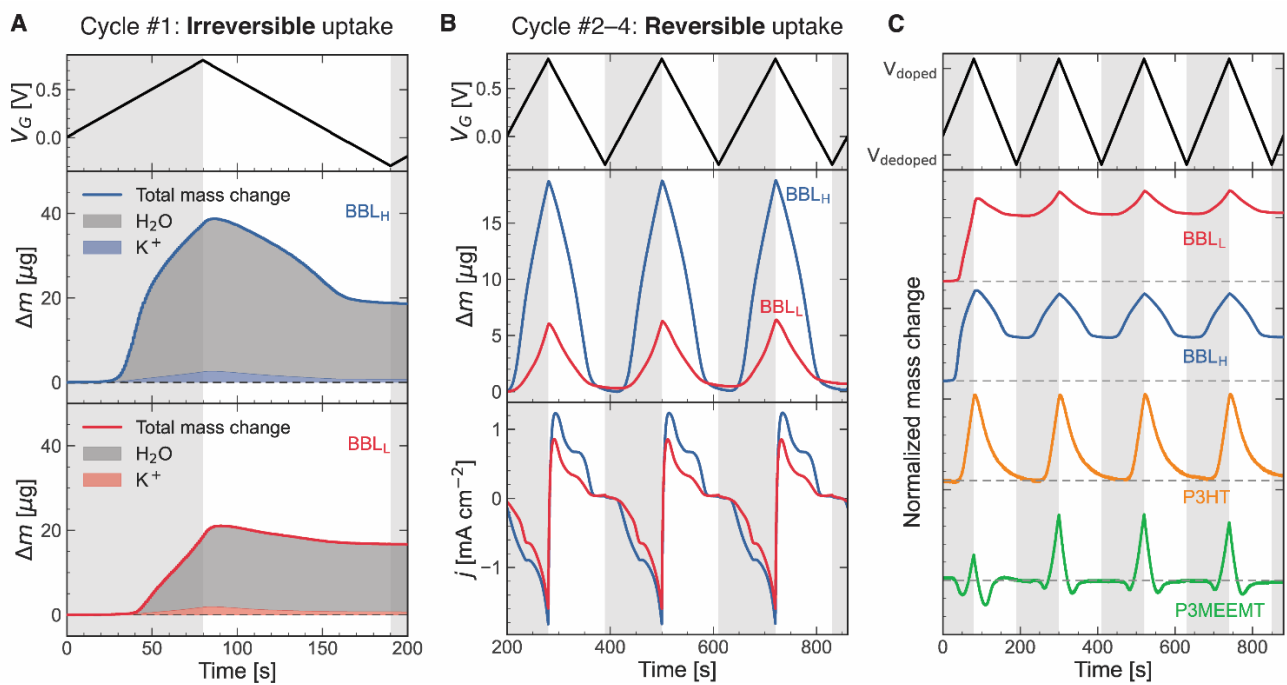

**Figure 2.** A) The *irreversible* total mass uptake measured by QCM during the *first* cycle of CV sweep (solid lines), the mass of potassium cations estimated from charge accumulated in the film (colored area), and the difference attributed to water uptake (grey area). B) The *reversible* mass changes measured by gravimetry during subsequent doping/dedoping cycles. Top: Voltage profile of scan cycles. Scan rate: 10 mV/s; the shadowed sections indicate the doping process. Middle: Mass changes of both BBL films. Bottom: Current density during CV cycles. C) The normalized mass change of **BBL$_L$**, **BBL$_H$**, P3HT and P3MEEMT over the first 4 doping/dedoping cycles. Each polymer mass change was normalized to its maximum mass change and offset by 1.1.

Furthermore, we investigate the same process in two other conjugated polymers, poly(3-hexylthiophene-2,5-diyl) (P3HT), and poly(3-thiophene-2,5-diyl) (P3MEEMT). (**Figure 2C**) We choose P3HT and P3MEEMT as example to exemplify the typical swelling behavior: P3HT has been reported as an OECT material without hydrophilic side-chains in chaotropic anion electrolytes,[16, 32] while P3MEEMT is a derivative with the same backbone with glycol side-chains which exhibits appreciable passive swelling.[17, 32, 55] In contrast to BBLs, neither P3HT nor P3MEEMT shows an irreversible mass retention after its first doping/dedoping cycle, even though maximum mass uptake of P3MEEMT in the first cycle is slightly different from others.

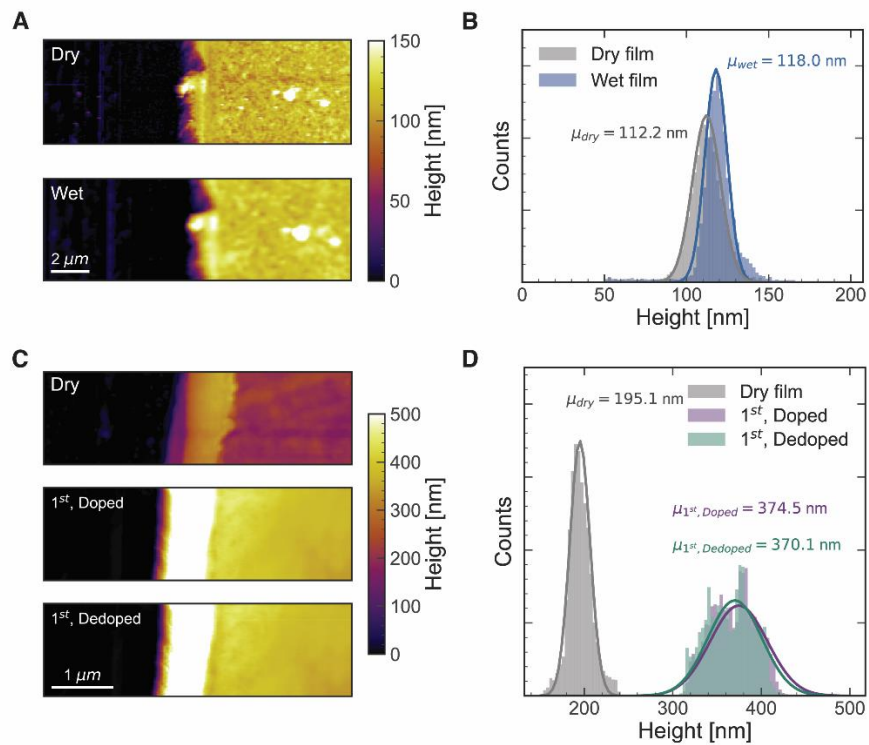

**Figure 3.** A) AFM height images and B) height histogram of the **BBL$_H$** dry film (top) and wet film (bottom) of the same area at the scratch edge on Au substrates (left side) C) AFM height images and D) height histogram of the **BBL$_H$** dry film (top), *in-operando* doped film (middle, 500 mV), and *in-operando* dedoped film (bottom, −300 mV) of the same area during the first doping/dedoping cycle at the scratch edge on Au substrates (left side) The dry films were measured in air. The wet and *in-operando* doped/dedoped films were measured in 100 mmol/L KCl electrolyte. All height profiles were fitted to Gaussian distributions.

Thus far, we have shown that both the commercial lower molecular mass and in-house synthesized higher molecular mass BBL polymers facilitate a fast, efficient bulk electrochemical doping along with a drastic ion/water mass uptake. Additionally, both BBLs undergo a film hydration step driven by the first electrochemical doping cycle that is irreversible upon subsequent doping cycles but that is reversible upon drying the films. To investigate the morphology and structural changes that accompany the hydration process of BBL, we used both AFM-based techniques and GIWAXS measurement on the high-performing, high molecular mass **BBL$_H$**. First, we measured the film thickness changes with *in-operando* AFM upon *passive* and *active* swelling. **Figure 3A–B** show the AFM images and height histograms of the pristine film before and after exposure to the electrolyte solution, without external bias. As probed via tapping mode AFM on the **BBL$_H$** under liquid, we observe negligible (≈5 %) swelling. This result of minimal swelling agrees with the previous QCM data showing a small, if any, mass change. However, on the other hand, **Figure 3C–D** show that when the doping potential (500 mV) is first applied during the initial cycle, the film swells over ≈90 %, and becomes rougher. After dedoping the film back to the neutral state at − 300 mV, the height of the film only decreases minimally by a few nanometers. Based on our previous eQCM result, we believe that the dramatic thickness increase observed here is a result of the initial film hydration during the first cycle, and the minimal shrink corresponds to the reversible active swelling.

Next, we investigate nanoscale morphology changes during active swelling using the AFM-based technique known as electrochemical strain microscopy (ESM). ESM has been used for probing nanoscale swelling due to the ion uptake in systems ranging from Li-ion battery materials[56-58] to conjugated polymers.[3, 16] We have previously applied it to reveal the spatial anticorrelation between ion uptake and the local crystallinity in p-type materials.[3] In ESM, an AC voltage is applied between the conductive tip and the polymer film in an electrolyte solution. Ion uptake due to the reduction/oxidation results in a local swelling, which is monitored at the contact resonance of the tip-sample interface. Because the frequency of the applied AC drive voltage is usually much faster than the full doping time for films, the observed local swelling under AC ESM is smaller than the swelling under DC conditions.[59-61] However, we can use ESM to probe the films under dynamic AC ion uptake at a range of different DC offset voltages.

**Figure 4A** and **Figure S16** show the ESM amplitude of the **BBL$_H$**, along with its topography image. Around the reduction onset potential, both films barely swell with an average amplitude of ≈500 pm. As we increase the tip bias, the ESM amplitude exhibits a larger swelling due to the ion/water injection into the film (**Figure 4B**). At the highest potential (300 mV over threshold), the **BBL$_H$** swells by an average amplitude of 5.83 nm (4.9 % swelling), which is a relatively large amount of AC swelling compared to other hydrophobic polymers like P3HT.[3] After doping at positive bias, then we scanned the DC tip bias offset to the dedoping condition and observed the amplitude decreasing back to a low level (< 1 %). This strong asymmetry in the ESM signal as a function of DC offset potential (i.e. swelling is observed only when the AC voltage takes the sample into a regime of electrochemical doping at one sign) confirms that the induced swelling is indeed the result of redox processes. Furthermore, we also operated the ESM measurements under different ionic concentrations (**Figure 4C** and **Figure S17**). As expected, the swelling tends to be greater (larger ESM amplitude) in a more concentrated electrolyte. The swelling over the film becomes large and more heterogenous with increasing bias, but the distribution of swelling volumes remains unimodal at all biases (**Figure 4B**). This behavior contrasts notably with our previous study on P3HT films, which showed heterogeneous swelling according to the local degree of crystallinity, and a clear bimodal distribution of swelling heights.[3] We interpret this result as a reflection of both more consistent uniformity in BBL films, and penetration of water and ions into the film on the nano- to molecular scales.

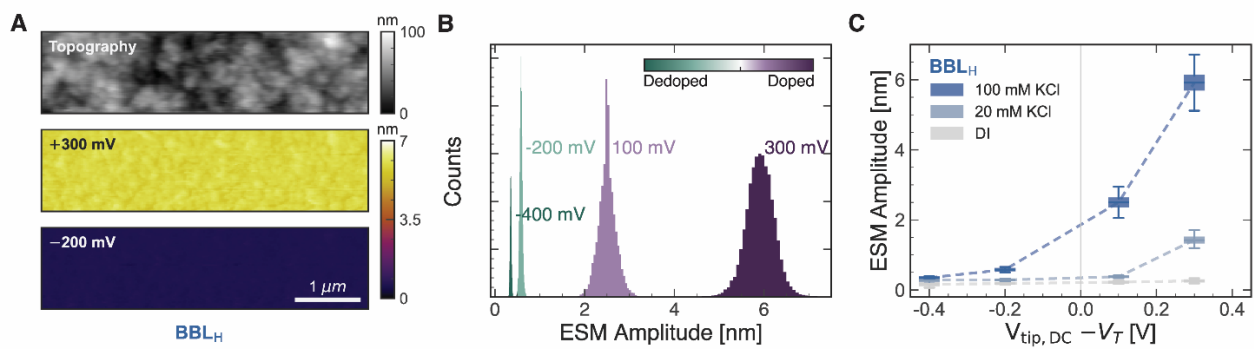

**Figure 4.** A) Topography and ESM amplitude images of the **BBL$_H$** films in 100 mmol/L KCl at a 500 mV AC drive voltage. ESM images were taken with the tip at different DC potentials over the threshold (as labeled). The measurements were taken in a random potential order to avoid artificial shifts of baseline. B) Histograms of the ESM amplitude of the same area under different bias conditions. C) The ESM amplitude of **BBL$_H$** as a function of $V_{DC} - V_T$ taken in different electrolyte concentrations. The dashed lines are used to guide the eye. The error bars represent one standard deviation of ESM amplitude distribution.

To probe molecular-level changes in film morphology upon ion-injection and water-uptake, we next carried out GIWAXS on **BBL$_H$** samples. Previous research by Inal and co-workers has explored the structural changes that occur upon doping with *ex-situ* GIWAXS.[35] However, Paulsen *et al.* and Flagg *et al.* have revealed that *ex-situ* GIWAXS, which probes the crystal structure under a dry film condition, may not capture the details of electrolyte swelling during operation.[62-64] Based on our previous eQCM data, we observed that BBL film undergoes a unique film swelling compared to typical conjugated polymers, which is exactly why electrochemical doping mechanism remains unresolved.

Thus, we investigate the crystalline structure changes using *in-operando* GIWAXS on **BBL$_H$**. We utilized a modified blade coater to apply a potential and draw electrolyte across the sample as described elsewhere.[64] In **Figure 5A**, we show the (100) lattice peaks as a function of electrolyte exposure and doping state. Before applying a bias, we first took the GIWAXS on the dry and wet BBL films after electrolyte exposure. The black trace ("dry") shows the initial out of plane (OOP) scattering of the dry film. Next, we performed one rolling drop passage while applying zero bias. As seen by the grey trace ("exposed"), exposure to the 100 mmol/L KCl does not change the lamellar spacing. However, upon application of a doping bias the lattice constant expands dramatically (≈31 %), from 8.3 Å to 10.9 Å. This expansion is consistent with both our eQCM and *in-operando* AFM measurements showing that significant film swelling only occurs upon doping, not due to initial electrolyte exposure. Continued cycling of the film between doped (purple traces) and dedoped (green traces) does modulate the lamellar spacing, where the doped form of the polymer has a more expanded lattice than the undoped form, which is consistent with other doped polymers.[64-65] Also, the lamellar spacing change between the doped and dedoped states in multiple cycles is reversible and reproducible (**Figure 5C**). Nevertheless, as long as the films remain wet, the (100) spacing never returns to the spacing of the initial exposed film, again consistent with the eQCM data which suggests irreversible water uptake on the first cycle. We note that the π–π lattice only minimally changes during this process (see **Figure S18** and **Figure S19**). Quantifying the swelling, we see the crystals only swelling ≈30 % volume by crystal expansion. However, we know the whole film is swelling ≈90 % in thickness – we therefore conclude that the majority of the swelling must be in the more amorphous regions of the BBL that are not scattering strongly. In contrast to *in-operando* GIWAXS, we also operated *ex-situ* GIWAXS on the **BBL$_H$**. (**Figure 5C**, also **Figure S20**) Obviously, the *ex-situ* GIWAXS significantly underestimates the lamellar expansion upon doping and fails to observe the ≈2.5 Å expansion due to water injection.

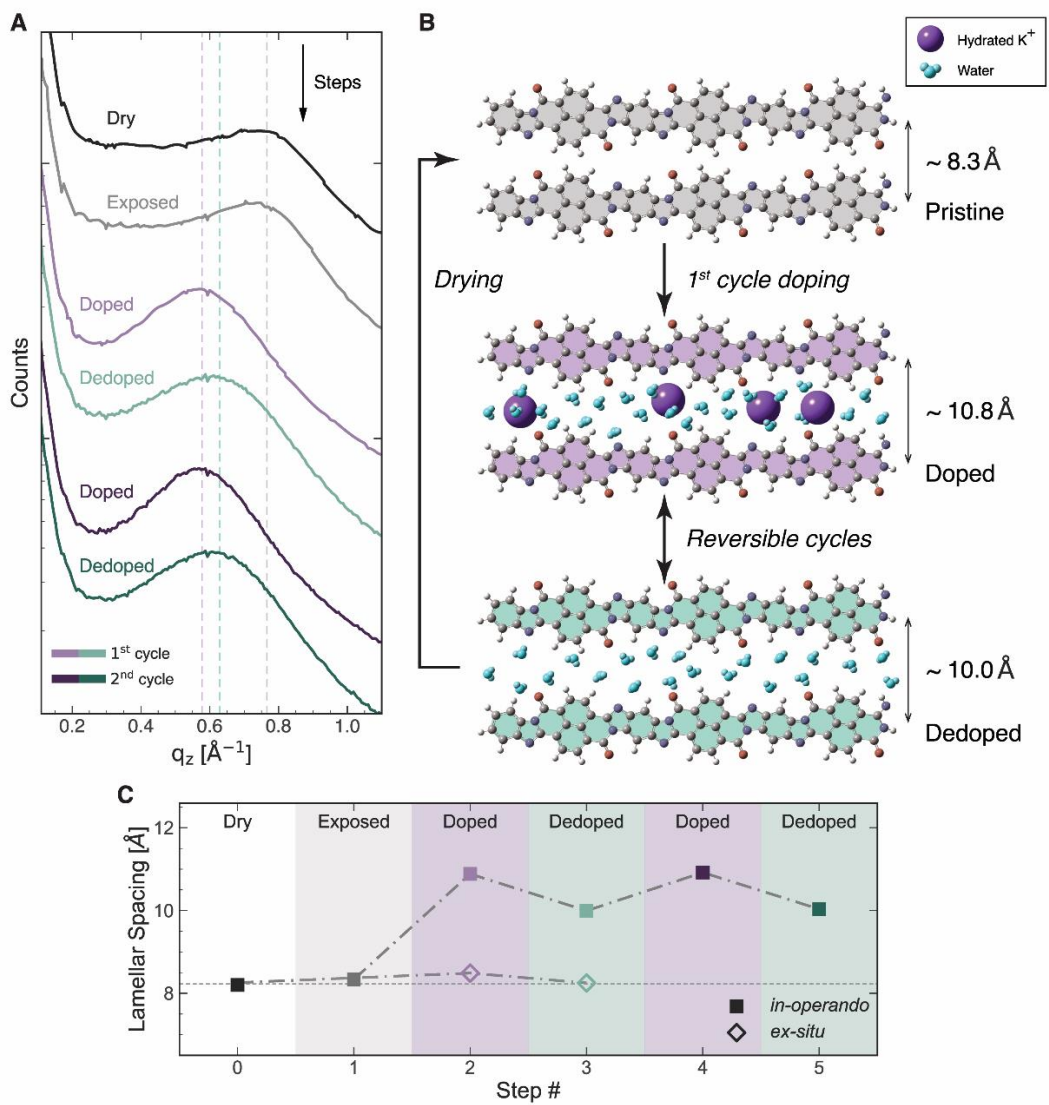

**Figure 5.** A) Out of plane (OOP) linecut profiles of *in-operando* GIWAXS patterns of **BBL$_H$** thin films under different film conditions. The dashed lines are used to guide the eye. B) Schematic showing the polymer hydration mechanism during the electrochemical doping for BBL. The purple spheres are hydrated K$^+$, and the cyan molecules are free waters. C) The (100) lamellar spacing changes along the *in-operando* (solid) and *ex-situ* (open) GIWAXS doping/dedoping cycle steps. Purple traces/markers: doped states; Green traces/markers: dedoped states.

**Table 2.** The d-spacings measured via *ex-situ* and *in-situ* GIWAXS. (Unit: Å)

|  | Pristine | Doped | Dedoped |
|---|---|---|---|
| ***In-plane (010)*** | | | |
| *Ex-situ* | 3.4 | 3.3 | 3.4 |
| *In-situ* | 3.4 | 3.3 | 3.3 |
| ***Out-of-plane (100)*** | | | |
| *Ex-situ* | 8.3 | 8.5 | 8.3 |
| *In-situ* | 8.3 | 10.9 | 10.0 |

Taking all these data together, we now propose a mechanism for how a hydrophobic ladder polymer like BBL performs as a fast operational OMIEC material (**Figure 5B**). Upon exposure to the electrolyte, the hydrophobic BBL is resistant to water uptake and undergoes little if any passive swelling. However, when the film is electrochemically doped under an applied potential for the first time, the counterions are accompanied by a large number of free water molecules and injected into both crystalline and amorphous domains in the film. Specifically in the crystalline regions, the lamellar spacing expands (≈30 %) much more than has been previously realized in BBL (≈15 %),[35] facilitating ions and water injection. When the film gets dedoped, the cations leave the films, along with some of the water molecules. However, a large amount of the water remains in the film, resulting in an irreversible mass gain, film thickness increase, and crystalline lamellar spacing expansion. This model suggests that due to the hydrophobic backbones and absence of side-chains, BBL film first needs to get hydrated with the help of the cation movement driven by an electric field. After the waters dragged in by cations, the carbonyl and amine groups along the backbones are sufficiently polar to retain the free water upon dedoping, which may not occur in a less polar backbone polymers. The remaining waters in the bulk film create a more hydrophilic microenvironment, facilitating a fast ion/water migration during the following doping/dedoping cycles. As a result, we see the switch ON or OFF time in milliseconds (**Table 1**). Furthermore, due to the rigid backbone of ladder polymer, even after ≈30 % lamellar expansion, the crystal structure is not significantly damaged such that the crystalline order is preserved on cycling to the dehydrated state after dedoping and drying, which enables good charge transport in a swelled film condition and good stability. To our knowledge, this is the first report of the unique film hydration activation process in a ladder polymer.

## CONCLUSIONS

We explore the electrochemical doping process of the side-chain-free n-type ladder polymer, BBL, comparing two different molecular mass versions: in-house synthesized **BBL$_H$** and commercial **BBL$_L$**. Like other conventional hydrophobic polymer films, BBL shows a minimal passive swelling when in contact with electrolytes. Surprisingly, without side-chains, BBL with a hydrophobic backbone facilitates cation injection along with significant water uptake during active swelling. This is attributed to a unique electrochemically irreversible film swelling during the first doping step, followed by a reversible swelling/deswelling accompanying the doping/dedoping cycles. With the combination of *in-situ* AFM, *ex-situ* and *in-operando* GIWAXS, we studied the detailed structural change upon these electrochemical redox reactions. We propose a model to account for the interesting swelling behavior. During the first doping step, BBL film gets hydrated with the help of ion movement and then maintains a hydrophilic environment due to the water retention in both amorphous and crystalline domains. The first hydration step of BBL helps explain why the hydrophobic BBL polymers could still allow ion injections during doping. Except for the first cycle, the mass uptake in the following cycles shows no retention. As a result, BBL retains advantages of good reversibility and connectivity due to the lack of large volume water-side-chain interactions. We also propose that some modest film swelling, separated from hydrated ion injection may be beneficial for overcoming the tradeoff of ionic and electronic transport for many applications ranging from biosensor to neuromorphic computing devices and thus may provide a design template for the synthesis of next generation OECT materials based on the lessons of BBL.

# ACKNOWLEDGMENTS

This paper is based on research supported primarily by the National Science Foundation, DMR–2003456. In-house synthesis of BBL was supported by the NSF (DMR-2003518). L.Q.F. acknowledges the support of a NIST-National Research Council fellowship. Part of this work was conducted at the Washington Nanofabrication Facility/Molecular Analysis Facility, a National Nanotechnology Coordinated Infrastructure (NNCI) site at the University of Washington with partial support from the National Science Foundation via awards NNCI-1542101 and NNCI-2025489. The *ex-situ* GIWAXS in this research used beamline 7.3.3 of the Advanced Light Source (ALS), which is a DOE Office of Science User Facility under contract no. DEAC02-05CH11231. The authors thank Chenhui Zhu and Eric Schaible at the ALS for assistance with GIWAXS data acquisition and analysis. The *in-situ* GIWAXS used beamline 11-BM (CMS) of the National Synchrotron Light Source, a U.S. Department of Energy (DOE) Office of Science User Facility operated for the DOE Office of Science by Brookhaven National Laboratory under contract no. DE-SC0012704. Certain commercial equipment, instruments, or materials are identified in this paper in order to specify the experimental procedure adequately. Such identification is not intended to imply recommendation or endorsement by the National Institute of Standards and Technology, nor is it intended to imply that the materials or equipment identified are necessarily the best available for the purpose.

# REFERENCES

1. Giridharagopal, R.; Guo, J.; Kong, J.; Ginger, D. S., Nanowire Architectures Improve Ion Uptake Kinetics in Conjugated Polymer Electrochemical Transistors. *ACS Appl. Mater. Interfaces* **2021,** *13* (29), 34616-34624.

2. Xu, K.; Ruoko, T. P.; Shokrani, M.; Scheunemann, D.; Abdalla, H.; Sun, H. D.; Yang, C. Y.; Puttisong, Y.; Kolhe, N. B.; Figueroa, J. S. M.; Pedersen, J. O.; Ederth, T.; Chen, W. M.; Berggren, M.; Jenekhe, S. A.; Fazzi, D.; Kemerink, M.; Fabiano, S., On the Origin of Seebeck Coefficient Inversion in Highly Doped Conducting Polymers. *Adv. Funct. Mater.* **2022,** *32* (20), 2112276.

3. Giridharagopal, R.; Flagg, L. Q.; Harrison, J. S.; Ziffer, M. E.; Onorato, J.; Luscombe, C. K.; Ginger, D. S., Electrochemical strain microscopy probes morphology-induced variations in ion uptake and performance in organic electrochemical transistors. *Nat. Mater.* **2017,** *16* (7), 737-742.

4. Szumska, A. A.; Maria, I. P.; Flagg, L. Q.; Savva, A.; Surgailis, J.; Paulsen, B. D.; Moia, D.; Chen, X.; Griggs, S.; Mefford, J. T.; Rashid, R. B.; Marks, A.; Inal, S.; Ginger, D. S.; Giovannitti, A.; Nelson, J., Reversible Electrochemical Charging of n-Type Conjugated Polymer Electrodes in Aqueous Electrolytes. *J. Am. Chem. Soc.* **2021,** *143* (36), 14795-14805.

5. Liu, H.; Yang, A.; Song, J.; Wang, N.; Lam, P.; Li, Y.; Law, H. K.; Yan, F., Ultrafast, sensitive, and portable detection of COVID-19 IgG using flexible organic electrochemical transistors. *Sci. Adv.* **2021,** *7* (38), eabg8387.

6. Inal, S.; Rivnay, J.; Suiu, A. O.; Malliaras, G. G.; McCulloch, I., Conjugated Polymers in Bioelectronics. *Acc. Chem. Res.* **2018,** *51* (6), 1368-1376.

7. Picca, R. A.; Manoli, K.; Macchia, E.; Sarcina, L.; Di Franco, C.; Cioffi, N.; Blasi, D.; Österbacka, R.; Torricelli, F.; Scamarcio, G.; Torsi, L., Ultimately Sensitive Organic Bioelectronic Transistor Sensors by Materials and Device Structure Design. *Adv. Funct. Mater.* **2019,** *30* (20), 1904513.

8. Bischak, C. G.; Flagg, L. Q.; Ginger, D. S., Ion Exchange Gels Allow Organic Electrochemical Transistor Operation with Hydrophobic Polymers in Aqueous Solution. *Adv. Mater.* **2020,** *32* (32), e2002610.

9. Mike, J. F.; Lutkenhaus, J. L., Recent advances in conjugated polymer energy storage. *Journal of Polymer Science Part B: Polymer Physics* **2013,** *51* (7), 468-480.

10. Volkov, A. V.; Sun, H. D.; Kroon, R.; Ruoko, T. P.; Che, C. Y.; Edberg, J.; Muller, C.; Fabiano, S.; Crispin, X., Asymmetric Aqueous Supercapacitor Based on p- and n-Type Conducting Polymers. *ACS Appl. Energy Mater.* **2019,** *2* (8), 5350-5355.

11. Sun, H.; Vagin, M.; Wang, S.; Crispin, X.; Forchheimer, R.; Berggren, M.; Fabiano, S., Complementary Logic Circuits Based on High-Performance n-Type Organic Electrochemical Transistors. *Adv. Mater.* **2018,** *30* (9), 1704916.

12. Andersson Ersman, P.; Lassnig, R.; Strandberg, J.; Tu, D.; Keshmiri, V.; Forchheimer, R.; Fabiano, S.; Gustafsson, G.; Berggren, M., All-printed large-scale integrated circuits based on organic electrochemical transistors. *Nat. Commun.* **2019,** *10* (1), 5053.

13. Keene, S. T.; Lubrano, C.; Kazemzadeh, S.; Melianas, A.; Tuchman, Y.; Polino, G.; Scognamiglio, P.; Cina, L.; Salleo, A.; van de Burgt, Y.; Santoro, F., A biohybrid synapse with neurotransmitter-mediated plasticity. *Nat. Mater.* **2020,** *19* (9), 969-973.

14. Fuller, E. J.; Keene, S. T.; Melianas, A.; Wang, Z.; Agarwal, S.; Li, Y.; Tuchman, Y.; James, C. D.; Marinella, M. J.; Yang, J. J.; Salleo, A.; Talin, A. A., Parallel programming of an ionic floating-gate memory array for scalable neuromorphic computing. *Science* **2019,** *364* (6440), 570-574.

15. Felder, D.; Femmer, R.; Bell, D.; Rall, D.; Pietzonka, D.; Henzler, S.; Linkhorst, J.; Wessling, M., Coupled Ionic–Electronic Charge Transport in Organic Neuromorphic Devices. *Adv. Theory Simul.* **2022,** *5* (6), 2100492.

16. Flagg, L. Q.; Giridharagopal, R.; Guo, J.; Ginger, D. S., Anion-Dependent Doping and Charge Transport in Organic Electrochemical Transistors. *Chem. Mater.* **2018,** *30* (15), 5380-5389.

17. Flagg, L. Q.; Bischak, C. G.; Quezada, R. J.; Onorato, J. W.; Luscombe, C. K.; Ginger, D. S., P-Type Electrochemical Doping Can Occur by Cation Expulsion in a High-Performing Polymer for Organic Electrochemical Transistors. *ACS Materials Lett.* **2020,** *2* (3), 254-260.

18. Kukhta, N. A.; Marks, A.; Luscombe, C. K., Molecular Design Strategies toward Improvement of Charge Injection and Ionic Conduction in Organic Mixed Ionic-Electronic Conductors for Organic Electrochemical Transistors. *Chem. Rev.* **2022,** *122* (4), 4325-4355.

19. Inal, S.; Malliaras, G. G.; Rivnay, J., Benchmarking organic mixed conductors for transistors. *Nat. Commun.* **2017,** *8* (1), 1767.

20. Moser, M.; Hidalgo, T. C.; Surgailis, J.; Gladisch, J.; Ghosh, S.; Sheelamanthula, R.; Thiburce, Q.; Giovannitti, A.; Salleo, A.; Gasparini, N.; Wadsworth, A.; Zozoulenko, I.; Berggren, M.; Stavrinidou, E.; Inal, S.; McCulloch, I., Side Chain Redistribution as a Strategy to Boost Organic Electrochemical Transistor Performance and Stability. *Adv. Mater.* **2020,** *32* (37), e2002748.

21. Li, P.; Lei, T., Molecular design strategies for high‐performance organic electrochemical transistors. *J. Polym Sci.* **2021,** *60* (3), 377-392.

22. Bischak, C. G.; Flagg, L. Q.; Yan, K.; Li, C. Z.; Ginger, D. S., Fullerene Active Layers for n-Type Organic Electrochemical Transistors. *ACS Appl. Mater. Interfaces* **2019,** *11* (31), 28138-28144.

23. Griggs, S.; Marks, A.; Bristow, H.; McCulloch, I., n-Type organic semiconducting polymers: stability limitations, design considerations and applications. *J. Mater. Chem. C Mater.* **2021,** *9* (26), 8099-8128.

24. Chen, X.; Marks, A.; Paulsen, B. D.; Wu, R.; Rashid, R. B.; Chen, H.; Alsufyani, M.; Rivnay, J.; McCulloch, I., n-Type Rigid Semiconducting Polymers Bearing Oligo(Ethylene Glycol) Side Chains for High-Performance Organic Electrochemical Transistors. *Angew. Chem. Int. Ed.* **2021,** *60* (17), 9368-9373.

25. Giovannitti, A.; Nielsen, C. B.; Sbircea, D. T.; Inal, S.; Donahue, M.; Niazi, M. R.; Hanifi, D. A.; Amassian, A.; Malliaras, G. G.; Rivnay, J.; McCulloch, I., N-type organic electrochemical transistors with stability in water. *Nat. Commun.* **2016,** *7*, 13066.

26. Sun, H. D.; Gerasimov, J.; Berggren, M.; Fabiano, S., n-Type organic electrochemical transistors: materials and challenges. *J. Mater. Chem. C* **2018,** *6* (44), 11778-11784.

27. Giovannitti, A.; Maria, I. P.; Hanifi, D.; Donahue, M. J.; Bryant, D.; Barth, K. J.; Makdah, B. E.; Savva, A.; Moia, D.; Zetek, M.; Barnes, P. R. F.; Reid, O. G.; Inal, S.; Rumbles, G.; Malliaras, G. G.; Nelson, J.; Rivnay, J.; McCulloch, I., The Role of the Side Chain on the Performance of N-type Conjugated Polymers in Aqueous Electrolytes. *Chem. Mater.* **2018,** *30* (9), 2945-2953.

28. Maria, I. P.; Paulsen, B. D.; Savva, A.; Ohayon, D.; Wu, R.; Hallani, R.; Basu, A.; Du, W.; Anthopoulos, T. D.; Inal, S.; Rivnay, J.; McCulloch, I.; Giovannitti, A., The Effect of Alkyl Spacers on the Mixed Ionic‐Electronic Conduction Properties of N‐Type Polymers. *Adv. Funct. Mater.* **2021,** *31* (14), 2008718.

29. Jeong, D.; Jo, I. Y.; Lee, S.; Kim, J. H.; Kim, Y.; Kim, D.; Reynolds, J. R.; Yoon, M. H.; Kim, B. J., High‐Performance n‐Type Organic Electrochemical Transistors Enabled by Aqueous Solution Processing of Amphiphilicity‐Driven Polymer Assembly. *Adv. Funct. Mater.* **2022,** *32* (16), 2111950.

30. Ohayon, D.; Savva, A.; Du, W.; Paulsen, B. D.; Uguz, I.; Ashraf, R. S.; Rivnay, J.; McCulloch, I.; Inal, S., Influence of Side Chains on the n-Type Organic Electrochemical Transistor Performance. *ACS Appl. Mater. Interfaces* **2021,** *13* (3), 4253-4266.

31. Feng, K.; Shan, W.; Ma, S.; Wu, Z.; Chen, J.; Guo, H.; Liu, B.; Wang, J.; Li, B.; Woo, H. Y.; Fabiano, S.; Huang, W.; Guo, X., Fused Bithiophene Imide Dimer-Based n-Type Polymers for High-Performance Organic Electrochemical Transistors. *Angew.Chem.* **2021,** *133*, 24400–24407.

32. Flagg, L. Q.; Bischak, C. G.; Onorato, J. W.; Rashid, R. B.; Luscombe, C. K.; Ginger, D. S., Polymer Crystallinity Controls Water Uptake in Glycol Side-Chain Polymer Organic Electrochemical Transistors. *J. Am. Chem. Soc.* **2019,** *141* (10), 4345-4354.

33. Savva, A.; Cendra, C.; Giugni, A.; Torre, B.; Surgailis, J.; Ohayon, D.; Giovannitti, A.; McCulloch, I.; Di Fabrizio, E.; Salleo, A.; Rivnay, J.; Inal, S., Influence of Water on the Performance of Organic Electrochemical Transistors. *Chem. Mater.* **2019,** *31* (3), 927-937.

34. Wu, H. Y.; Yang, C. Y.; Li, Q.; Kolhe, N. B.; Strakosas, X.; Stoeckel, M. A.; Wu, Z.; Jin, W.; Savvakis, M.; Kroon, R.; Tu, D.; Woo, H. Y.; Berggren, M.; Jenekhe, S. A.; Fabiano, S., Influence of Molecular Weight on the Organic Electrochemical Transistor Performance of Ladder-Type Conjugated Polymers. *Adv. Mater.* **2022,** *34* (4), e2106235.

35. Surgailis, J.; Savva, A.; Druet, V.; Paulsen, B. D.; Wu, R.; Hamidi‐Sakr, A.; Ohayon, D.; Nikiforidis, G.; Chen, X.; McCulloch, I.; Rivnay, J.; Inal, S., Mixed Conduction in an N‐Type Organic Semiconductor in the Absence of Hydrophilic Side‐Chains. *Adv. Funct. Mater.* **2021,** *31* (21), 2010165.

36. Proctor, C. M.; Rivnay, J.; Malliaras, G. G., Understanding volumetric capacitance in conducting polymers. *Journal of Polymer Science Part B: Polymer Physics* **2016,** *54* (15), 1433-1436.

37. Sahalianov, I.; Singh, S. K.; Tybrandt, K.; Berggren, M.; Zozoulenko, I., The intrinsic volumetric capacitance of conducting polymers: pseudo-capacitors or double-layer supercapacitors? *RSC Adv.* **2019,** *9* (72), 42498-42508.

38. Wang, S.; Sun, H.; Ail, U.; Vagin, M.; Persson, P. O.; Andreasen, J. W.; Thiel, W.; Berggren, M.; Crispin, X.; Fazzi, D.; Fabiano, S., Thermoelectric Properties of Solution-Processed n-Doped Ladder-Type Conducting Polymers. *Adv. Mater.* **2016,** *28* (48), 10764-10771.

39.     Gu, K.; Snyder, C. R.; Onorato, J.; Luscombe, C. K.; Bosse, A. W.; Loo, Y. L., Assessing the Huang-Brown Description of Tie Chains for Charge Transport in Conjugated Polymers. *ACS Macro Lett.* **2018,** *7* (11), 1333-1338.

40.     Wang, H.; Xu, Y.; Yu, X.; Xing, R.; Liu, J.; Han, Y., Structure and Morphology Control in Thin Films of Conjugated Polymers for an Improved Charge Transport. *Polymers* **2013,** *5* (4), 1272-1324.

41.     Noriega, R.; Rivnay, J.; Vandewal, K.; Koch, F. P.; Stingelin, N.; Smith, P.; Toney, M. F.; Salleo, A., A general relationship between disorder, aggregation and charge transport in conjugated polymers. *Nat. Mater.* **2013,** *12* (11), 1038-1044.

42.     Kline, R. J.; McGehee, M. D.; Kadnikova, E. N.; Liu, J.; Fréchet, J. M. J., Controlling the Field-Effect Mobility of Regioregular Polythiophene by Changing the Molecular Weight. *Adv. Mater.* **2003,** *15* (18), 1519-1522.

43.     Li, J.; Zhao, Y.; Tan, H. S.; Guo, Y.; Di, C. A.; Yu, G.; Liu, Y.; Lin, M.; Lim, S. H.; Zhou, Y.; Su, H.; Ong, B. S., A stable solution-processed polymer semiconductor with record high-mobility for printed transistors. *Sci. Rep.* **2012,** *2* (1), 754.

44.     Hayashi, S.; Yamamoto, S. I.; Koizumi, T., Effects of molecular weight on the optical and electrochemical properties of EDOT-based pi-conjugated polymers. *Sci. Rep.* **2017,** *7* (1), 1078.

45.     Meier, H.; Stalmach, U.; Kolshorn, H., Effective conjugation length and UV/vis spectra of oligomers. *Acta Polymer.* **1997,** *48* (9), 379-384.

46.     Gierschner, J.; Huang, Y. S.; Van Averbeke, B.; Cornil, J.; Friend, R. H.; Beljonne, D., Excitonic versus electronic couplings in molecular assemblies: The importance of non-nearest neighbor interactions. *J. Chem. Phys.* **2009,** *130* (4), 044105.

47.     Vagin, M.; Gueskine, V.; Mitraka, E.; Wang, S.; Singh, A.; Zozoulenko, I.; Berggren, M.; Fabiano, S.; Crispin, X., Negatively‐Doped Conducting Polymers for Oxygen Reduction Reaction. *Adv. Energy Mater.* **2021,** *11* (3), 2002664.

48.     Nicolini, T.; Surgailis, J.; Savva, A.; Scaccabarozzi, A. D.; Nakar, R.; Thuau, D.; Wantz, G.; Richter, L. J.; Dautel, O.; Hadziioannou, G.; Stingelin, N., A Low-Swelling Polymeric Mixed Conductor Operating in Aqueous Electrolytes. *Adv. Mater.* **2021,** *33* (2), e2005723.

49.     Denison, D. R., Linearity of a Heavily Loaded Quartz Crystal Microbalance. *Journal of Vacuum Science and Technology* **1973,** *10* (1), 126-129.

50.     Behrndt, K. H., Long-Term Operation of Crystal Oscillators in Thin-Film Deposition. *Journal of Vacuum Science and Technology* **1971,** *8* (5), 622-626.

51.     Vogt, B. D.; Lin, E. K.; Wu, W.-L.; White, C. C., Effect of Film Thickness on the Validity of the Sauerbrey Equation for Hydrated Polyelectrolyte Films. *J. Phys. Chem. B* **2004,** *108* (34), 12685-12690.

52.     Cendra, C.; Giovannitti, A.; Savva, A.; Venkatraman, V.; McCulloch, I.; Salleo, A.; Inal, S.; Rivnay, J., Role of the Anion on the Transport and Structure of Organic Mixed Conductors. *Adv. Funct. Mater.* **2019,** *29* (5), 1807034.

53.     Mancinelli, R.; Botti, A.; Bruni, F.; Ricci, M. A.; Soper, A. K., Hydration of sodium, potassium, and chloride ions in solution and the concept of structure maker/breaker. *J. Phys. Chem. B* **2007,** *111* (48), 13570-13577.

54. Mahler, J.; Persson, I., A study of the hydration of the alkali metal ions in aqueous solution. *Inorg. Chem.* **2012,** *51* (1), 425-438.

55. Schmode, P.; Savva, A.; Kahl, R.; Ohayon, D.; Meichsner, F.; Dolynchuk, O.; Thurn-Albrecht, T.; Inal, S.; Thelakkat, M., The Key Role of Side Chain Linkage in Structure Formation and Mixed Conduction of Ethylene Glycol Substituted Polythiophenes. *ACS Appl. Mater. Interfaces* **2020,** *12* (11), 13029-13039.

56. Balke, N.; Kalnaus, S.; Dudney, N. J.; Daniel, C.; Jesse, S.; Kalinin, S. V., Local detection of activation energy for ionic transport in lithium cobalt oxide. *Nano Lett.* **2012,** *12* (7), 3399-3403.

57. Balke, N.; Jesse, S.; Kim, Y.; Adamczyk, L.; Tselev, A.; Ivanov, I. N.; Dudney, N. J.; Kalinin, S. V., Real space mapping of Li-ion transport in amorphous Si anodes with nanometer resolution. *Nano Lett.* **2010,** *10* (9), 3420-3425.

58. Balke, N.; Jesse, S.; Morozovska, A. N.; Eliseev, E.; Chung, D. W.; Kim, Y.; Adamczyk, L.; Garcia, R. E.; Dudney, N.; Kalinin, S. V., Nanoscale mapping of ion diffusion in a lithium-ion battery cathode. *Nat. Nanotech.* **2010,** *5* (10), 749-754.

59. Balke, N.; Jesse, S.; Kim, Y.; Adamczyk, L.; Ivanov, I. N.; Dudney, N. J.; Kalinin, S. V., Decoupling electrochemical reaction and diffusion processes in ionically-conductive solids on the nanometer scale. *ACS Nano.* **2010,** *4* (12), 7349-7357.

60. Alikin, D. O.; Romanyuk, K. N.; Slautin, B. N.; Rosato, D.; Shur, V. Y.; Kholkin, A. L., Quantitative characterization of the ionic mobility and concentration in Li-battery cathodes via low frequency electrochemical strain microscopy. *Nanoscale* **2018,** *10* (5), 2503-2511.

61. Morozovska, A. N.; Eliseev, E. A.; Balke, N.; Kalinin, S. V., Local probing of ionic diffusion by electrochemical strain microscopy: Spatial resolution and signal formation mechanisms. *J. Appl. Phys.* **2010,** *108* (5), 053712.

62. Paulsen, B. D.; Giovannitti, A.; Wu, R.; Strzalka, J.; Zhang, Q.; Rivnay, J.; Takacs, C. J., Electrochemistry of Thin Films with In Situ/Operando Grazing Incidence X-Ray Scattering: Bypassing Electrolyte Scattering for High Fidelity Time Resolved Studies. *Small* **2021,** *17* (42), e2103213.

63. Paulsen, B. D.; Wu, R.; Takacs, C. J.; Steinruck, H. G.; Strzalka, J.; Zhang, Q.; Toney, M. F.; Rivnay, J., Time-Resolved Structural Kinetics of an Organic Mixed Ionic-Electronic Conductor. *Adv. Mater.* **2020,** *32* (40), e2003404.

64. Flagg, L. Q.; Asselta, L. E.; D'Antona, N.; Nicolini, T.; Stingelin, N.; Onorato, J. W.; Luscombe, C. K.; Li, R.; Richter, L. J., In Situ Studies of the Swelling by an Electrolyte in Electrochemical Doping of Ethylene Glycol-Substituted Polythiophene. *ACS Appl. Mater. Interfaces* **2022,** *14* (25), 29052-29060.

65. Bischak, C. G.; Flagg, L. Q.; Yan, K.; Rehman, T.; Davies, D. W.; Quezada, R. J.; Onorato, J. W.; Luscombe, C. K.; Diao, Y.; Li, C. Z.; Ginger, D. S., A Reversible Structural Phase Transition by Electrochemically-Driven Ion Injection into a Conjugated Polymer. *J. Am. Chem. Soc.* **2020,** *142* (16), 7434-7442.